\documentclass[12pt]{article}

\usepackage{epsfig,latexsym,amsfonts,amsmath,amsthm,amssymb,amsbsy,multirow,slashed,wasysym,textcomp,subfigure,hyperref,wrapfig,datetime,comment,mathtools,cancel,cite,mathrsfs}
\usepackage{datetime2}
\usepackage[font={footnotesize,sl},bf]{caption}

\def\ee{\end{equation}}
\def\bea{\begin{eqnarray}}
\def\eea{\end{eqnarray}}
\newcommand{\beq}{\begin{eqnarray}}
\newcommand{\eqq}{\end{eqnarray}}
 \newcommand{\badat}{\begin{alignedat}}
 \newcommand{\eadat}{\end{alignedat}}

\newcommand{\eal}[1]{\be \begin{aligned} #1 \end{aligned}\end{equation}} 
\newcommand{\eqn}[1]{\be #1 \end{equation}} 
\newcommand{\eqa}[1]{\bea  #1\end{eqnarray}}

\newcommand{\mO}{\mathcal{O}}

\newcommand{\mY}{\mathcal{Y}}

\newcommand{\bz}{\bar{z}}

\newcommand{\outstate}{\langle \text{out}|}
\newcommand{\instate}{|\text{in}\rangle}
\newcommand{\scri}{\mathscr{I}}

\long\def\new#1\endnew{{\bf #1}}		
\long\def\del#1\enddel{}

\def\del{\partial}

\usepackage{xcolor}
\definecolor{oldmauve}{rgb}{0.4, 0.19, 0.28}
\definecolor{pansypurple}{rgb}{0.47, 0.09, 0.29}
\definecolor{burgundy}{rgb}{0.5, 0.0, 0.13}
\definecolor{carminepink}{rgb}{0.92, 0.3, 0.26}
\definecolor{blue(pigment)}{rgb}{0.2, 0.2, 0.6}
\definecolor{darkseagreen}{rgb}{0.56, 0.74, 0.56}
\definecolor{darkspringgreen}{rgb}{0.09, 0.45, 0.27}
\definecolor{ceruleanblue}{rgb}{0.16, 0.32, 0.75}

\newcommand{\be}{\begin{eqnarray}}
\newcommand{\en}{\end{eqnarray}}

\def\bz{{\bar z}}

\author{}
\numberwithin{equation}{section} 

\begin{document}

\begin{titlepage}
  \thispagestyle{empty}

  \begin{center}  
  
\vspace*{2cm}
  
{\LARGE\textbf{Logarithmic soft graviton theorems from}}
\vskip0.5cm
{\LARGE\textbf{superrotation Ward identities}}

\vskip1cm
Shreyansh Agrawal$^\star$\footnote{\fontsize{8pt}{10pt}\selectfont \ \href{mailto:sagrawal@sissa.it}{sagrawal@sissa.it}},
Laura Donnay$^\star$\footnote{\fontsize{8pt}{10pt}\selectfont \ \href{mailto:ldonnay@sissa.it}{ldonnay@sissa.it}}, Kevin Nguyen$^\dagger$\footnote{\fontsize{8pt}{10pt}\selectfont\  \href{mailto:kevin.nguyen@kcl.ac.uk}{kevin.nguyen@kcl.ac.uk}}, Romain Ruzziconi$^*$\footnote{\fontsize{8pt}{10pt}\selectfont\ \href{mailto:Romain.Ruzziconi@maths.ox.ac.uk}{romain.ruzziconi@maths.ox.ac.uk}}

\vskip0.5cm

\normalsize
\medskip

$^\star$\textit{SISSA,
Via Bonomea 265, 34136 Trieste, Italy}

\textit{INFN, Sezione di Trieste,
Via Valerio 2, 34127, Italy \\
\vspace{2mm}
}

$^\dagger$\textit{Department of Mathematics, King's College London,\\
The Strand, London WC2R 2LS, UK\\
\vspace{2mm}
} 

$^*$\textit{Mathematical Institute, University of Oxford, \\ Andrew Wiles Building, Radcliffe Observatory Quarter, \\
Woodstock Road, Oxford, OX2 6GG, UK}

\end{center}

\vskip0.5cm

\begin{abstract}
Soft graviton theorems receive one-loop contributions that are logarithmic in the energy of the soft graviton, and which are closely related to tails of gravitational waveforms. We demonstrate that these logarithmic corrections are encoded in the Ward identity of superrotation symmetries, i.e.~they follow from conservation of superrotation charge across spatial infinity~$i^0$. Our proof relies on a careful analysis of the radiative phase space admitting such gravitational tails, and the determination of the fluxes through null infinity~$\scri$ that act as canonical generators of superrotations on both gravitational and matter fields. All logarithmic terms are derived from the fluxes through correlations of the supertranslation Goldstone mode, provided care is taken in manipulating gravitationally interacting (i.e.~dressed) rather than free fields. In cases where massive particles take part in the scattering process, logarithmic corrections also partly arise from the superrotation charge generator at timelike infinity~$i^\pm$.    
\end{abstract}

\end{titlepage}

\tableofcontents
\section{Introduction}

Soft graviton theorems and their relation with asymptotic symmetries play an important role in developing an understanding of the infrared structure of quantum gravity. Influential work \cite{Strominger:2013jfa,He:2014laa,Adamo:2014yya,Kapec:2014opa} has indeed shown that conservation of supertranslation charge is equivalent to Weinberg's leading soft graviton theorem \cite{Weinberg:1965nx}, while conservation of superrotation charge encodes the tree-level subleading soft graviton theorem \cite{Cachazo:2014fwa} (the case of massive particle scattering was treated in \cite{Campiglia:2015kxa}). The soft graviton theorems are therefore the most direct physical consequence of the extended BMS symmetries \cite{Bondi:1962px,Sachs:1962wk,Sachs:1962zza,Barnich:2009se,Barnich:2010eb} in the context of scattering amplitudes. 

Soft graviton theorems receive loop corrections beyond the leading order $O(\omega^{-1})$ in the soft energy $\omega$ expansion, and it is of primary interest to assess whether these also consistently follow from BMS conservation laws, i.e., whether BMS symmetries hold in the quantum theory. However the literature on loop corrections to soft graviton theorems contains seemingly different results. On the one hand a general analysis of Bern, Davies and Nohle showed that loop corrections appear at order $O(\omega^0)$ in the soft expansion \cite{Bern:2014oka}, while more recently Sen et al. \cite{Laddha:2018myi,Laddha:2018vbn,Sahoo:2018lxl,Saha:2019tub} ~have obtained corrections at order $O(\ln \omega)$ (see also \cite{Ciafaloni:2018uwe} for a derivation of log terms in the eikonal approach) and related them to gravitational tail effects in classical gravity \cite{Blanchet:1987wq,Blanchet:1993ec}. In previous works \cite{Donnay:2022hkf,Pasterski:2022djr},  a detailed derivation of the $O(\omega^0)$ corrections as a consequence of the conservation of superrotation charge was provided (see also \cite{He:2017fsb,He:2023lvk}). Establishing this correspondence required significant control over the radiative gravitational phase space, and therefore informs us about foundational aspects of the theory. The goal of the present work will be to unify these earlier results with the $O(\ln \omega)$ corrections found by Sahoo and Sen \cite{Sahoo:2018lxl} (see also the recent generalisation \cite{Krishna:2023fxg}). 

Soft graviton theorems associated with superrotations also play a central role in flat holography. Indeed in that context the subleading soft graviton theorem maps onto the conformal Ward identity of a two-dimensional celestial CFT \cite{Kapec:2016jld}, or that of a three-dimensional Carrollian CFT \cite{Donnay:2022aba,Donnay:2022wvx}. The stress tensor of the corresponding boundary theory is identified with the integral density of the soft superrotation flux, whose determination is thus of central importance.

The paper is organised as follows. In section~\ref{section: log corrections} we review the $O(\ln \omega)$ corrections found by Sahoo and Sen \cite{Sahoo:2018lxl} and observe that they coincide with the $O(\omega^0)$ corrections worked out in \cite{Bern:2014oka} for the case where all scattered particles are massless, provided one substitutes $\ln \omega$ by the cutoff used to regulate infrared divergences. This strongly suggests that the methods developed in previous works \cite{Donnay:2022hkf,Pasterski:2022djr} and which accounted for the $O(\omega^0)$ corrections to the subleading soft graviton theorem can also account for the generic $O(\ln \omega)$ corrections. To demonstrate this, our main task will be to extend our previous analysis to the case where scattered particles can be massive. To this end we review in section~\ref{sec: 4} the exponentiation of infrared divergence in scattering amplitudes without soft graviton emission \cite{Weinberg:1965nx}, and their description in terms of correlation functions of the supertranslation Goldstone mode \cite{Himwich:2020rro,Arkani-Hamed:2020gyp,Nguyen:2021ydb}. We point out that a slight modification of the Goldstone two-point function needs to be considered in order to account for the so-called `Coulomb' phases corresponding to classical long-range interaction between asymptotic states. In section~\ref{sec: 2} we describe the radiative phase space of asymptotically flat gravity and its organisation with respect to extended BMS symmetries. In particular, we provide a symplectic form such that these symmetries are realised canonically and identify the canonical generators with BMS fluxes \cite{Donnay:2021wrk}. This generalises the construction provided in \cite{Donnay:2022hkf} and building upon \cite{Ashtekar:1981bq,He:2014laa,Campiglia:2021bap,Donnay:2021wrk} to the case where gravitational tails are allowed in the phase space. This results in a new term in the soft superrotation flux compared to previous work \cite{Donnay:2021wrk,Donnay:2022hkf}, which however vanishes in a superrotation frame of reference or in case gravitational tails need not be considered explicitly.  Finally in section~\ref{sec: 5} we derive all the logarithmic corrections found by Sahoo and Sen \cite{Sahoo:2018lxl} from superrotation charge conservation. While some of these corrections directly arise from the soft superrotation flux as in \cite{Donnay:2022hkf}, for massive particle scattering some corrections also arise from the generator of superrotation at timelike infinities $i^\pm$ when considering matter fields interacting with the gravitational field, i.e.~dressed fields rather than free fields.\footnote{Loop corrections in QED were similarly discussed in \cite{Campiglia:2019wxe}.}

\section{Logarithmic corrections to soft graviton theorems}
\label{section: log corrections}
Soft graviton theorems relate a scattering amplitude $\mathcal{M}_{n}(p_1,\dots,p_n)$ of $n$ hard particles of momenta $p_i$ to the amplitude $\mathcal{M}_{n+1}(q,p_1,\dots,p_n)$ that contains an additional external soft graviton of momentum $q=\omega \hat q$. Assuming a power series expansion in the soft momentum, the tree-level soft graviton theorem at leading and subleading order in $\omega$ takes the form  \cite{Weinberg:1965nx,Cachazo:2014fwa}
\begin{equation}
\mathcal{M}_{n+1} \stackrel{\omega \to 0}{=}\left[\omega^{-1}\, \hat S^{(0)}_n+S^{(1)}_n\right]\mathcal{M}_n+O(\omega)\,,
\end{equation}
where the leading and subleading soft factors are respectively given by 
\begin{equation}
\label{soft_factors}
\hat S^{(0)}_n=\frac{\kappa}{2} \sum_{i=1}^n \frac{p_i^\mu p_i^\nu \varepsilon_{\mu\nu}(\hat q)}{p_i \cdot \hat{q}}\,, \qquad S^{(1)}_n=-\frac{i \kappa}{2} \sum_{i=1}^n \frac{p_i^\mu\,  \varepsilon_{\mu\nu}(\hat q)\, q_\lambda}{p_i \cdot q}\, \left( J_i^{\lambda\nu}+S_i^{\lambda \nu} \right)\equiv S_n^{(1)J}+S_n^{(1)S}\,,
\end{equation}
where $\kappa=\sqrt{32\pi G}$, $\varepsilon_{\mu\nu}(\hat q)$ is the polarisation tensor of the soft graviton, $J_i^{\mu \nu}=-i(p_i^\mu \partial_{p_i}^\nu-p_i^\nu \partial_{p_i}^\mu)$ and $S_i^{\mu\nu}$ are the orbital and spin angular momentum of the $i$-th particle, respectively.

Sahoo and Sen have shown that one-loop corrections generate logarithmic corrections which actually dominate over the $O(\omega^0)$ term in the expansion \cite{Sahoo:2018lxl},
\begin{equation}
\label{log theorem}
\mathcal{M}_{n+1} \stackrel{\omega \to 0}{=}\left[\omega^{-1}\,  \hat S^{(0)}_n-\frac{\kappa^2}{4}\ln \omega    \,S^{(\textrm{ln})}_n\right]\mathcal{M}_n+O(\omega^0)\,.
\end{equation}
Note that these loop corrections necessarily introduce one (or more) infrared length scale $R$ so as to make the argument of the logarithm $\ln (\omega R)$ dimensionless. Here we will restrict our attention to the $\ln \omega$ dependence and absorb the $\ln R$ contributions into the $O(\omega^0)$ terms in the soft expansion. These logarithmic corrections are explicitly given by
\begin{align}
\nonumber
S^{\textrm{(ln)}}_n&=\frac{i\kappa}{8\pi} \sum_i\frac{\varepsilon_{\mu \nu}p^\mu_i p^\nu_i}{p_i\cdot q} \sum_{j} \delta_{\eta, \eta_j}\, q\cdot p_j\\
\nonumber
& +\frac{i\kappa}{16\pi}\sum_i\frac{\varepsilon_{\mu \nu}p^\nu_i q_\rho}{p_i\cdot q} \sum_{j\neq i} \delta_{\eta_i,\eta_j} (p_i\cdot p_j)\,(p^\mu_i p^\rho_j-p^\mu_jp^\rho_i)\frac{2(p_i\cdot p_j)^2-3 p_i^2p_j^2}{\left[(p_i\cdot p_j)^2-p^2_i p^2_j\right]^{3/2}}\\
\label{Sahoo Sen log}
&-\frac{\kappa}{8\pi^2} \sum_i\frac{\varepsilon_{\mu \nu}p^\nu_i p^\nu_i}{p_i\cdot q}\sum_{j} q\cdot p_j \ln |\hat q \cdot \hat p_j|  \\
\nonumber
&-\frac{\kappa}{32\pi^2}\sum_i\frac{p^\mu_i\varepsilon_{\mu \nu} q_\lambda}{p_i\cdot q}\left(p^\lambda_i \frac{\partial}{\partial p_{i\nu}}-p^\nu_i \frac{\partial}{\partial p_{i\lambda}}\right)  \sum_{j\neq i} \frac{2(p_i\cdot p_j)^2-p_i^2 p_j^2}{[(p_i\cdot p_j)^2-p_i^2 p_j^2]^{1/2}} \, \ln\left(\frac{p_i\cdot p_j+\sqrt{(p_i\cdot p_j)^2-p_i^2 p_j^2}}{p_i\cdot p_j-\sqrt{(p_i\cdot p_j)^2-p_i^2 p_j^2}}\right)\,.
\end{align}
Note that the formulae of Sahoo and Sen are presented with $\kappa=2$ in our conventions. In \cite{Sahoo:2018lxl} the first two lines and last two lines are referred to as the `classical' and `quantum' contributions, respectively, even though they all arise from one-loop diagrams. The reason for this distinction is that the `classical' terms can be obtained from purely classical considerations. In particular they naturally map to late-time gravitational tails \cite{Laddha:2018vbn,Saha:2019tub}. 
The first term in \eqref{Sahoo Sen log} represents the effect of gravitational drag on the soft graviton due to the other hard particles in the final state. The second term represents the effect of late time gravitational radiation due to the late time acceleration of the particles via long range gravitational interaction.
The `quantum terms' were obtained by explicit evaluation of one-loop diagrams in a theory of minimally coupled scalars in \cite{Sahoo:2018lxl}, and the generalisation to spinning fields was recently given in \cite{Krishna:2023fxg}. In particular the first term comes from a region of loop integration where the loop momentum is small compared to $\omega$ but large compared to the infrared cut-off, while the second term comes from regions where the loop momentum is large compared to $\omega$ but small compared to all other energy scales involved in the process.

There is a significantly more compact way to write \eqref{Sahoo Sen log}, which will turn out very useful in carrying our analysis. For this we introduce the relative velocity of the particles $i$ and $j$,
\begin{equation}
\label{betaij}
\beta_{ij}\equiv \sqrt{1-(\hat p_i \cdot \hat p_j)^{-2}}\,,
\end{equation}
where $\hat p_i$ is the normalised momentum satisfying $\hat p_i^2=-1$, and the quantities 
\begin{equation}
\label{Sigma}
\sigma_n=\frac{1}{2(8\pi)^2}\sum_{i j} \eta_i \eta_j\, m_i m_j\,  \frac{1+\beta_{ij}^2}{\beta_{ij}\sqrt{1-\beta_{ij}^2}}  \left(i\pi \delta_{\eta_i,\eta_j}-\frac{1}{2} \ln \frac{1+\beta_{ij}}{1-\beta_{ij}} \right)\,,
\end{equation}
\begin{equation}
\label{Sigma prime definition}
\hat \sigma'_{n+1}(\hat q)=  \frac{1}{2(4\pi)^2} \sum_{i=1}^n (p_i \cdot \hat q)  \ln ( \hat p_i \cdot \hat q)\,. 
\end{equation}
In the next section, $\sigma_n$ will be related with Weinberg's infrared divergent factor \eqref{weinfact}. Accordingly, the sum in \eqref{Sigma} should be understood as including $i=j$ terms for the real part only (see e.g. \cite{Yennie:1961ad,Weinberg:1965nx,Alessio:2024wmz}). 
In writing \eqref{Sigma} we have assumed that all hard particles are massive. If the $i$-th particle is massless, then the corresponding terms in the sum are replaced by
\begin{equation}
\frac{1}{(8\pi)^2}\, \eta_i \eta_j\, |p_i \cdot p_j| \left(i\pi \delta_{\eta_i, \eta_j}-\ln | \hat p_i \cdot \hat p_j|\right)=\frac{1}{(8\pi)^2}\, (p_i \cdot p_j) \ln (\hat p_i \cdot \hat p_j)\,,
\end{equation}
which is valid whether the $j$-th particle is massless or massive. In fact $\hat \sigma'_{n+1}(\hat q)$ derives from $\sigma_{n+1}$ in case where the $(n+1)$-th particle is a (soft) graviton with normalised momentum $\hat q$,
\begin{equation}
\sigma_{n+1}=\sigma_n+\hat \sigma'_{n+1}(\hat q)\,.
\end{equation}
In terms of these quantities, the logarithmic soft factor \eqref{Sahoo Sen log} can be simply written\footnote{Such a formula can be found in (6.29) of \cite{Sahoo:2018lxl}, with $\widehat S^{(1)}_{\text{gr}}=S^{(1)J}_n$. Notice that, while $\sigma_n$ includes real $i=j$ terms, these terms simply reduce to $m_i^2$ (times a numerical factor). Therefore, these $i=j$ terms drop out when acting with the subleading soft operator and do not appear in $S^{(\ln)}$.}
\begin{equation}
\label{log factor rewritten}
S^{(\ln)}_n(\hat q)=-8\left( \hat \sigma'_{n+1}(\hat q)\, \hat S^{(0)}_n(\hat q)-S^{(1)J}_n(\hat q)\, \sigma_n\right)\,,    
\end{equation}
where $S^{(1)J}_n(\hat q)$ is the orbital part of the subleading soft operator defined in \eqref{soft_factors}, and where we have kept the dependence on the soft graviton momentum direction $\hat q$ explicit. The fact that the logarithmic corrections do not depend on the spin of the particles has been confirmed in \cite{Krishna:2023fxg} . 

At this point it is very informative to make a comparison between the expression \eqref{log factor rewritten} and the earlier result of Bern, Davies and Nohle \cite{Bern:2014oka} regarding infrared divergent one-loop corrections to the soft graviton theorem in purely massless scatterings, namely
\begin{equation}
\label{Bern}
\mathcal{M}_{n+1} \stackrel{\omega \to 0}{=}\left[\omega^{-1}\, \hat S^{(0)}_n+S^{(1)}_n+ \kappa^2\, \epsilon^{-1}\left(  \hat \sigma_{n+1}'\, \hat S^{(0)}_n  - \left(S^{(1)J}_n \sigma_n\right) \right)\right]\mathcal{M}_n+\mO(\omega)\,,
\end{equation}
where $\epsilon=(4-d)/2$ is an infrared regulator in dimensional regularisation. There is an obvious correspondence between the infrared divergent terms in \eqref{Bern} and the logarithmic terms derived by Sahoo and Sen and given in \eqref{log factor rewritten}, provided one makes the replacement 
\begin{equation}
\label{epsilon log omega identification}
\epsilon^{-1} \quad \longleftrightarrow \quad \ln \omega\,.
\end{equation}
We can understand this correspondence in the following way. If instead of dimensional regularisation we used energy cutoffs as Weinberg originally discussed \cite{Weinberg:1965nx}, we would replace $\epsilon$ by
\begin{equation}
\epsilon^{-1} \quad \longleftrightarrow \quad \ln \frac{\Lambda}{\lambda}\,,
\end{equation}
where $\lambda$ is the infrared energy regulator while $\Lambda$ is a dividing energy scale below which virtual gravitons are considered soft and some approximations can be applied. We see that we recover the identification \eqref{epsilon log omega identification} provided we use the energy $\omega$ of the emitted soft graviton 
as the dividing energy scale $\Lambda$. The use of $\omega$ as an upper cutoff in one-loop integrals actually appears explicitly in \cite{Sahoo:2018lxl}.

\section{Soft factorisation and supertranslation Goldstone}
\label{sec: 4}
An important feature of gravitational scattering is that infrared divergences factorise from the rest of the amplitude \cite{Weinberg:1965nx}. For a scattering of $n$ hard particles, this factorisation takes the form
\begin{equation}
\label{soft factorisation}
\mathcal{M}_n=\mathcal{M}_{\text{soft}}\, \mathcal{M}_{\text{finite}}\,,
\end{equation}
where $\mathcal{M}_{\text{soft}}$ accounts for the infrared divergences coming from soft virtual gravitons propagating between external legs,
\begin{equation}
\label{weinfact}
\mathcal{M}_{\text{soft}}=\exp\left[\frac{1}{\epsilon} \frac{\kappa^2}{2(8\pi)^2}\sum_{i j} \eta_i \eta_j\, m_i m_j\,  \frac{1+\beta_{ij}^2}{\beta_{ij}\sqrt{1-\beta_{ij}^2}}  \left(i\pi \delta_{\eta_i,\eta_j}-\frac{1}{2} \ln \frac{1+\beta_{ij}}{1-\beta_{ij}} \right)\right]\,,
\end{equation}
with $\beta_{ij}$ the relative velocity of particle $i$ and $j$ given in \eqref{betaij}. Notice also the relationship $\mathcal{M}_{\text{soft}}=\exp[\kappa^2 \epsilon^{-1} \sigma_n]$. The soft factor \eqref{weinfact} actually vanishes as the infrared regulator is removed $(\epsilon \to 0^+)$ but fortunately drops out of the experimentally observable inclusive cross-sections. Note that the imaginary part of the exponent in \eqref{weinfact} only involves pairs of particles that are both incoming or outgoing, and corresponds to the effect of the gravitational potential attraction between these particles. Such pure phases are often referred to as `Coulomb phases' by abuse of language and in analogy with quantum electrodynamics.

In the language of celestial holography, the scattering amplitude $\mathcal{M}_n$ is interpreted as the correlation function of $n$ `celestial' operators\footnote{Usually the term `celestial' is reserved to operators recast in boost basis, while here we deal with momentum basis operators.},
\begin{equation}
\mathcal{M}_n=\outstate \mathcal{S} \instate=\langle \mathcal O_1\, ...\, \mathcal  O_n \rangle\,,
\end{equation}
where each operator $\mathcal  O_i$ represents an external particle with momentum $p_i$. The factorisation \eqref{weinfact} of the amplitudes motivates the factorisation of the celestial operators themselves \cite{Himwich:2020rro},
\begin{equation}
\label{factorisation operators}
\mathcal  O_1(p_1)=\mathcal{W}_1(p_1)\, \tilde {\mathcal O}_1(p_1)\,,
\end{equation}
such that $\mathcal{M}_{\text{soft}}$ and $\mathcal{M}_{\text{finite}}$ be respectively obtained as
\begin{equation}
\label{Msoft celestial}
\mathcal{M}_{\text{soft}}=\langle \mathcal{W}_1\, ...\, \mathcal{W}_n \rangle\,, \qquad  \mathcal{M}_{\text{finite}}=\langle \Tilde{\mathcal O}_1\, ...\, \Tilde{\mathcal  O}_n \rangle\,.
\end{equation}
The $\tilde {\mathcal O}(p)$'s are dressed operators and $\mathcal{M}_{\text{finite}}$ is the scattering amplitude of dressed particles.

In practice a specific parametrisation of massless momenta $q$ in terms of energy and coordinates on the celestial sphere is used, 
\begin{equation}
\label{massless para}
q^\mu = \omega\, \eta\, (1+z\bar z, z+\bar z, -i(z-\bar z), 1-z\bar z) \equiv \omega\, \hat q^\mu\,,
\end{equation}
while for a massive momentum $p$ a parametrisation in terms of coordinates $(\rho,z,\bar z)$ on a three-dimensional hyperboloid is used,
\begin{equation}
\label{massive para}
p^\mu = \frac{m}{2\rho}\, \eta\, (1+\rho^2(1+z\bar z),\rho^2(z+\bar z), -i\rho^2 (z-\bar z), -1+\rho^2 (1-z \bar z)) \equiv m\, \hat p^\mu \,.
\end{equation}
We will reserve the letter $q$ for massless momenta only, while the letter $p$ will always refer to the momentum of one of the hard external particles that can be either massive or massless.

The operators $\tilde {\mathcal O}_i$ are to be understood as (conformally) dressed operators and the vertex operators $\mathcal{W}_i$ as the corresponding operator dressings \cite{Himwich:2020rro,Arkani-Hamed:2020gyp,Pasterski:2021dqe,Nguyen:2023ibj}. For a massless particle with momentum $p=\omega\, \hat p$ parametrised as in \eqref{massless para}, the vertex operator is given in terms of the supertranslation Goldstone mode $C^{(0)}$ by \cite{Himwich:2020rro}
\begin{equation}
\label{massless W}
\mathcal{W}(p)=e^{i \omega C^{(0)}(\hat p)}\,, 
\end{equation}
while for a massive particle with momentum $p=m\, \hat p$ parametrised as in \eqref{massive para}, it is given by
\begin{equation}
\label{massive W}
\mathcal{W} (p) = \exp \left[\frac{im}{2} \int d^2 \hat q\, \mathcal{G}(\hat p;\hat q)\, C^{(0)}(\hat q) \right]\,, 
\end{equation}
where it is understood that $\hat q=\hat q(w,\bar w)$ shares the time-orientation of $\hat p$, namely $\eta(\hat q)=\eta(\hat p)$, and where $d^2 \hat q=dw\, d\bar w$ and $\mathcal G(\hat p;\hat q)$ is the `bulk-boundary propagator' \cite{Campiglia:2015lxa}
\begin{equation}
\mathcal G(\hat p;\hat q)=\frac{1}{\pi} \left(\frac{\rho}{1+\rho^2|z-w|^2} \right)^3\,.
\end{equation}
Using the vertex operators \eqref{massless W}-\eqref{massive W} together with the Goldstone two-point function, 
\begin{equation}
\label{Goldstone correlator}
\begin{split}
\langle C^{(0)}(\hat q_i) C^{(0)}(\hat q_j) \rangle&=-\frac{1}{\epsilon} \frac{\kappa^2}{2(4\pi)^2}\, (\hat q_i \cdot \hat q_j) \ln (\hat q_i \cdot \hat q_j)\\
&=\frac{1}{\epsilon} \frac{\kappa^2}{(4\pi)^2}\, \eta_i \eta_j\,  |z_{ij}|^2 \left(\ln |z_{ij}|^2 -i \pi \delta_{\eta_i,\eta_j} \right)\,,
\end{split}
\end{equation} 
it was shown that the soft factor \eqref{weinfact} is indeed recovered \cite{Himwich:2020rro}. Note that the imaginary part of the above expression had not been explicitly considered in previous work \cite{Himwich:2020rro,Nguyen:2021ydb,Donnay:2022hkf}, although it is necessary here in order to account for the `Coulomb phases' in \eqref{weinfact}. This new contribution will actually play a crucial role in recovering the classical logarithmic corrections to the soft theorems. Also note that the Goldstone two-point function is still a Green's function for the operator $\partial^2 \bar \partial^2$ in agreement with the effective action derived in \cite{Nguyen:2021ydb}, whereas the new imaginary term corresponds to a different choice of boundary condition.

With this at hand, the insertion of $C^{(0)}$ in the $\mathcal S$-matrix can be computed along the same line as in \cite{Donnay:2022hkf}, by considering the soft factor of a scattering with an additional massless particle of momentum $q=\omega\, \hat q(z,\bar z)$ and taking suitable derivative with respect to $\omega$,  
\begin{equation}
\begin{split}
\langle C^{(0)}(\hat q) \mathcal{W}_1\, ...\,  \mathcal{W}_n \rangle&=-i \partial_\omega \langle \mathcal{W}(q) \mathcal{W}_1\, ...\,  \mathcal{W}_n  \rangle \big|_{\omega=0}=-i \partial_\omega \mathcal{M}_{\text{soft}}(q,p_1,...,p_n) \big|_{\omega=0}\\
&=-i \frac{\kappa^2}{\epsilon} \hat \sigma'_{n+1}(\hat q)\, \langle \mathcal{W}_1\, ...\,  \mathcal{W}_n \rangle\,.
\end{split}
\end{equation}
Since the operators $\tilde {\mathcal O}_i$ and $\mathcal{W}_i$ do not interact, this property is also valid at the level of the full $\mathcal{S}$-matrix itself,
\begin{equation}
\label{Goldstone insertion}
\outstate C^{(0)}(\hat q)\, \mathcal{S} \instate=-i \frac{\kappa^2}{\epsilon}\, \hat \sigma'_{n+1}(\hat q)\, \outstate \mathcal{S} \instate\,.
\end{equation}
This relation complements the leading soft graviton theorem, which describes the insertion of the leading soft news $\mathcal{N}^{(0)}_{zz}$ (see below \eqref{soft news}) into $\mathcal{S}$-matrix elements \cite{He:2014laa},
\begin{equation}
\label{N0 insertion}
\outstate\, \mathcal{N}^{(0)}_{zz}(\hat q)\, \mathcal{S} \instate=-\frac{\kappa}{16\pi} \hat S^{(0)+}_n(\hat q)\, \outstate \mathcal{S} \instate\,.
\end{equation}

\section{Radiative phase space at null infinity}
\label{sec: 2}
In this section we give a detailed description of the radiative phase space at null infinity $\mathscr{I}^+$, following notations and conventions of \cite{Donnay:2022hkf}. We extend that previous analysis by allowing for gravitational tails as they directly relate to the logarithmic corrections to the soft graviton theorems which we aim to describe in this work. 

We cover $\mathscr{I}^+$ with coordinates $(u,z, \bar{z})$, where $u$ is a retarded time and $(z, \bar{z})$ are complex stereographic coordinates on the celestial sphere. We choose the representative of the degenerate boundary metric to be $ds^2|_{\mathscr{I}^+} = 0\, du^2 + 2\, dz\, d\bar{z}$ and we denote $\partial\equiv \partial_z$, $\bar  \partial\equiv \partial_\bz$. The radiative data is encoded in the asymptotic shear $C_{zz}(u,z,\bar{z})$ ($C_{zz}^* = C_{\bar z \bar z}$) and the Bondi news tensor $N_{zz} = \partial_u C_{zz}$. We will assume the following falloffs in $u$ as $u\to \pm \infty$, which are compatible with the action of BMS symmetries on the phase space \cite{Compere:2018ylh,Compere:2020lrt}, and are sufficiently weak to encompass gravitational tails \cite{Blanchet:1987wq,Blanchet:1993ec,Blanchet:2020ngx}, 
\begin{equation}
     C_{zz} = (u + C_\pm) N_{zz}^{vac} - 2 \partial^2 C_\pm + \frac{1}{u} C^{L, \pm}_{zz} + o(u^{-1})\,, \quad N_{zz} = N_{zz}^{vac} - \frac{C_{zz}^{L,\pm}}{u^2} +  o(u^{-2}) \, .
    \label{falloff in u}
\end{equation} Here, $C_{\pm}(z, \bar z)$ correspond to the values of the supertranslation field at the corners $\mathscr{I}^+_\pm$ of null infinity and encoding the displacement memory effect \cite{Strominger:2014pwa}. The presence of the new term $C^{L, \pm}_{zz}(z, \bar{z})$ in the expansion \eqref{falloff in u} is necessary in order to account for gravitational tails \cite{Laddha:2018myi,Laddha:2018vbn,Sahoo:2018lxl,Saha:2019tub}. The vacuum news tensor $N_{zz}^{vac}(z)$ \cite{Compere:2016jwb,Compere:2018ylh}, identified with the tracefree part of the Geroch tensor \cite{Geroch1977,Campiglia:2020qvc,Nguyen:2022zgs}, is given in terms of a Liouville field $\varphi (z)$,
\begin{equation}
    N_{zz}^{vac} = \frac{1}{2}(\partial\varphi)^2 - \partial^2 \varphi \,.
    \label{Liouville stress tensor}
\end{equation}
The latter encodes the refraction/velocity kick memory effects \cite{Compere:2018ylh}. To make the covariance of the expressions under the superrotations manifest, it is useful to introduce the derivative operators~\cite{Barnich:2021dta,Campiglia:2020qvc,Donnay:2021wrk,Freidel:2021ytz}
\begin{equation}
    \begin{split}
       \mathscr{D}\phi_{h, \bar{h}} = [\partial - h \partial \varphi] \phi_{h, \bar{h}}\,,  \qquad 
       \bar{\mathscr{D}}\phi_{h, \bar{h}} =  [\bar{\partial} - \bar{h} \bar{\partial} \bar{\varphi}] \phi_{h, \bar{h}}\,,
    \end{split}  \label{derivative operators conformal}
\end{equation} which, when acting on conformal fields $\phi_{h,\bar{h}}$, produce conformal fields of weights $(h+1, \bar{h})$ and $(h, \bar{h}+1)$, respectively. 

The radiative data can be split into hard and soft variables \cite{Campiglia:2020qvc,Campiglia:2021bap,Donnay:2022hkf}. To do so we define $C^{(0)}_{zz}$, $\tilde{C}_{zz}$ and $\tilde{N}_{zz}$ through
\begin{equation}
\label{defs fields}
 \begin{split} 
C_{zz}&= u N^{vac}_{zz} + C^{(0)}_{zz} + \tilde C_{zz}\,, \qquad N_{zz}= N_{zz}^{vac} + \tilde{N}_{zz}\,, \\
C^{(0)}_{zz}&=-2 \mathscr{D}^2 C^{(0)}\,, \qquad \qquad \qquad C^{(0)}= \frac{1}{2}(C_++C_-)\,.
 \end{split} 
\end{equation} 
Moreover, the leading soft news $\mathcal{N}^{(0)}_{zz}$ and subleading soft news $\mathcal{N}^{(1)}_{zz}$ are defined by
\begin{equation}
    \mathcal{N}^{(0)}_{zz} \equiv \int_{-\infty}^{+\infty} du \, \tilde{N}_{zz} = -4 \mathscr{D}^2 N^{(0)}\,, \qquad \mathcal{N}^{(1)}_{zz} \equiv \int_{-\infty}^{+\infty} du \,  u \tilde{N}_{zz} \,.
    \label{soft news}
\end{equation} 

Now we wish to provide a symplectic form on this radiative phase space such that extended BMS symmetries are canonical transformations, and identify the corresponding canonical generators. The hard variables are $\{\tilde{C}_{zz}\,, \tilde{N}_{zz} \}$ while the soft variables are $\{C^{(0)}\,, \varphi\,, \mathcal{N}^{(0)}_{zz}\,, \mathcal{N}^{(1)}_{zz} \}$. The appropriate symplectic form $\Omega$ on the radiative phase space at $\mathscr{I}^+$ is the sum of a hard and a soft contribution \cite{Campiglia:2021bap,Donnay:2022hkf},
\begin{equation}
\begin{split}
    &\Omega =  \Omega^{hard} + \Omega^{soft} \,,\\
    &\Omega^{hard} = \frac{1}{32\pi G} \int_{\mathscr{I}^+} du\, d^2 z \left[ \delta \tilde{N}_{zz} \wedge \delta \tilde{C}_{\bar{z}\bar{z}}  + c.c.\right] + \text{matter sector}\,, \\
    &\Omega^{soft} = \frac{1}{32\pi G} \int_{\mathcal{S}}  d^2 z \left[\delta \mathcal{N}^{(0)}_{zz} \wedge \delta C^{(0)}_{\bar{z}\bar{z}}  + \delta \Pi_{zz}  \wedge \delta N_{\bar{z}\bar{z}}^{vac} +c.c.   \right] \,,
\end{split}
\label{symplectic structure full}
\end{equation} with the constraint
\begin{equation}
    \Pi_{zz} = 2 \mathcal{N}^{(1)}_{zz} + C^{(0)} \mathcal{N}^{(0)}_{zz} + (\varphi + \bar{\varphi} ) \Delta C^L_{zz}\,.
    \label{Pi variable}
\end{equation} 
Notice the crucial shift\footnote{$\Pi_{zz}^{here} = \Pi_{zz}^{there} +(\varphi + \bar{\varphi} ) \Delta C^L_{zz}$.} of $\Pi_{zz}$ compared to \cite{Campiglia:2021bap,Donnay:2022hkf} which involves $\Delta C^L_{zz} \equiv C^{L,+}_{zz} - C^{L,-}_{zz}$. Under BMS transformations parametrised by supertranslations $\mathcal{T}(z, \bar{z})$ and superrotations $(\mathcal{Y}(z), \bar{\mathcal{Y}}(\bar z))$, we have
\begin{equation}
\begin{split}
    &\delta_{(\mathcal{T},\mathcal{Y}, \bar{\mathcal{Y}})} \Delta C^L_{zz} = (\mathcal{Y} \partial + \bar{\mathcal{Y}} \bar{\partial} + \partial \mathcal{Y} - \bar{\partial} \bar{\mathcal{Y}} ) \Delta C^L_{zz}\,, \\
    &\delta_{(\mathcal{T},\mathcal{Y}, \bar{\mathcal{Y}})} \mathcal{N}^{(1)}_{zz} = (\mathcal{Y} \partial + \bar{\mathcal{Y}} \bar{\partial} + \partial \mathcal{Y} - \bar{\partial} \bar{\mathcal{Y}} )  \mathcal{N}^{(1)}_{zz} - \frac{1}{2} (\partial \mathcal{Y} + \bar{\partial} \bar{\mathcal{Y}} )  \Delta C^L_{zz} - \mathcal{T} \mathcal{N}^{(0)}_{zz}\,.
\end{split} \label{transformation log pair}
\end{equation} 
Using \eqref{transformation log pair}, one can now show that the combination \eqref{Pi variable} transforms nicely under BMS transformations,
\begin{equation}
    \delta_{(\mathcal{T},\mathcal{Y}, \bar{\mathcal{Y}})} \Pi_{zz} = (\mathcal{Y} \partial + \bar{\mathcal{Y}} \bar{\partial} + \partial \mathcal{Y} - \bar{\partial} \bar{\mathcal{Y}} ) \Pi_{zz} - \mathcal{T} \mathcal{N}^{(0)}_{zz}\,.
\end{equation} 
The transformation of the other canonical variables remain exactly the same as in \cite{Donnay:2022hkf} and will not be repeated here.

We then obtain the canonical generators of BMS symmetries, also known as BMS fluxes, by contracting the symplectic structure \eqref{symplectic structure full} with the corresponding Hamiltonian vector fields, 
\begin{equation}
    i_{\delta_{(\mathcal{T}, \mathcal{Y}, \bar{\mathcal{Y}})}} \Omega^{soft} = \delta F^{soft}_{(\mathcal{T}, \mathcal{Y}, \bar{\mathcal{Y}})}\,, \qquad i_{\delta_{(\mathcal{T}, \mathcal{Y}, \bar{\mathcal{Y}})}} \Omega^{hard} = \delta F^{hard}_{(\mathcal{T}, \mathcal{Y}, \bar{\mathcal{Y}})}\,.
    \label{canonical generators}
\end{equation} 
The supertranslation fluxes are given by
\begin{equation}
\begin{split}
     &F_{\mathcal{T}}^{hard} = - \frac{1}{16\pi G} \int du d^2z\, \mathcal{T} \left[ \tilde{N}_{zz}  \tilde{N}_{\bar{z}\bar{z}}+16 \pi G\, T_{uu}^{(2)} \right] \,,\\
     &F_{\mathcal{T}}^{soft} = \frac{1}{8\pi G} \int d^2 z\, \mathcal{T} \left[ \mathscr{D}^2\mathcal{N}^{(0)}_{\bar{z}\bar{z}} \right]\,,
\end{split} \label{supertranslation fluxes}
\end{equation} where $T_{\mu\nu}^{(2)}$ denotes the order $O(r^{-2})$ in the expansion of the $\mu\nu$-component of the matter stress tensor,
while the superrotation fluxes are given by
\begin{equation}
\label{superrotation fluxes}
\begin{split}
     &F_{\mathcal{Y}}^{hard} = \frac{1}{16\pi G} \int du d^2 z\,  {\mathcal{Y}} \left[\frac{3}{2}\tilde C_{zz} \partial \tilde N_{\bar z \bar z}+ \frac{1}{2}\tilde N_{\bar z \bar z} \partial \tilde C_{zz}+\frac{u}{2} \partial (\tilde N_{zz} \tilde N_{\bar z \bar z})+16\pi G \left(\frac{u}{2}\partial_zT_{uu}^{(2)}-T_{uz}^{(2)}\right)\right] \,,\\
     &F_{\mathcal{Y}}^{soft} = \frac{1}{16\pi G}\int d^2 z\,  \mathcal{Y}  \left[-{\mathscr D}^3 \left( \mathcal N_{\bar z \bar z}^{(1)} + \frac{1}{2} (\varphi + \bar{\varphi} ) \Delta C^L_{\bar z \bar z} \right) + \frac{3}{2} C^{(0)}_{zz} \mathscr{D} \mathcal{N}^{(0)}_{\bar z \bar z}+ \frac{1}{2}\mathcal{N}^{(0)}_{\bar z \bar z} \mathscr{D} C_{zz}^{(0)}  \right]\,,
\end{split}
\end{equation} 
together with the complex conjugate expressions associated with $\bar{\mathcal{Y}}$. These fluxes are identical to those presented in \cite{Donnay:2022hkf} when setting $\Delta C^L_{\bar z \bar z} = 0$. Note that $F_{\mathcal{Y}}^{soft}$ contains logarithmic infrared divergences through $\mathcal N_{\bar z \bar z}^{(1)}$, in agreement with earlier observations \cite{Donnay:2022hkf,Compere:2023qoa}. 

For completeness we express these fluxes as a difference of charges between the corners $\scri^+_\pm$ of future null infinity. Following the notations and conventions of \cite{Donnay:2022hkf}, and introducing the mass and angular momentum aspects in terms of Newman--Penrose scalars,\footnote{Notice that the falloffs \eqref{falloff in u} imply a logarithmic divergence in the angular momentum aspect ($\mathcal{N} \sim \ln u$) \cite{Compere:2023qoa}, although the mass aspect remains finite ($\mathcal{M} \sim u^{0}$). This is the same divergence as that appearing in the flux \eqref{superrotation fluxes} through $\mathcal{N}^{(1)}_{\bar z \bar z}$.}
\begin{equation}
   \mathcal{M} = -\frac{1}{2}(\Psi^0_2 + \bar{\Psi}^0_2) \,, \qquad \mathcal{N} = -\Psi^0_1 + u\, \bar{\partial}\Psi^0_2 \,,
    \label{momenta in NP}
\end{equation}
the total BMS fluxes can be written
\begin{equation}
F(\scri^+)\equiv F^{hard}+F^{soft}=Q(\scri^+_+)-Q(\scri^+_-)\,,
\end{equation}
with
\begin{equation}
Q(\scri^+_\pm) = \frac{1}{32\pi G} \int  d^2z  \left[8\mathcal{T} \mathcal{M} + \mathcal{Y} \left[4 \bar{\mathcal{N}}-{\partial} (N_{zz}^{vac} C^L_{\bar{z}\bar{z}}) +{\mathscr D}^3 ( (\varphi + \bar{\varphi} )  C^L_{\bar z \bar z} )\right]+ c.c. \right]\Big|_{\scri^+_\pm}\,. \label{charge at corner}
\end{equation} 
The last two terms featuring $C^L_{\bar z \bar z}$ give new contributions compared to the expression of BMS charges given in \cite{Donnay:2022hkf}. 

\section{Superrotation Ward identity}
\label{sec: 5}
We finally come to the derivation of the logarithmic corrections to soft graviton theorems from the Ward identity of superrotations. First we briefly remind the general procedure allowing the derivation of soft graviton theorems from BMS conservation laws \cite{Strominger:2013jfa,He:2014laa,Kapec:2014opa,Campiglia:2015kxa}. The key ingredient is the conservation of the BMS charges across spatial infinity $i^0$,
\begin{equation}
\label{BMS conservation}
Q(\scri^+_-)=Q(\scri^-_+)\equiv Q(i^0)\,.
\end{equation}
It should be pointed out that genuine conservation laws at spatial infinity have only been established for the global BMS group containing supertranslations but no superrotations \cite{Compere:2011ve}. Alternatively \eqref{BMS conservation} can be established provided gravitational fields satisfy suitable antipodal matching relations between $\scri^+_-$ and $\scri^-_+$ in the asymptotic solution space of interest. This in particular requires to connect gravitational solutions in asymptotic gauges adapted to $\scri^\pm$ and $i^0$, respectively, which has become an active subject of investigations \cite{Friedrich:2002ru,Troessaert:2017jcm,Henneaux:2018cst,Mohamed:2021rfg,Prabhu:2019fsp,Prabhu:2021cgk,Capone:2022gme,Compere:2023qoa}. For superrotations the relevant antipodal matching is that of the angular momentum aspect, which has been established in an asymptotic solution space without gravitational tails near spatial infinity, i.e., in a phase space where the physical news tensor falls-off as $\tilde N_{zz}=o(|u|^{-2})$ in the limit $u \to -\infty$ \cite{Capone:2022gme}. The logarithmic corrections to the soft theorems discussed here are however tightly connected to existence of such tails \cite{Laddha:2018vbn,Saha:2019tub}, and to establish the validity of the antipodal matching in this context therefore constitutes an important open problem. In what follows we will simply assume the validity of \eqref{BMS conservation} and work out its implications. We also work in a reference superrotation frame where $\varphi=\bar \varphi=0$ and therefore the derivative operator \eqref{derivative operators conformal} is the standard derivative $\mathscr{D}=\partial_z\equiv\partial$.

Writing the BMS fluxes at past and future null infinity as a net difference of charges,
\begin{equation}
\begin{split}
F(\scri^+)&=Q(\scri^+_+)-Q(\scri^+_-)\,,\\
F(\scri^-)&=Q(\scri^-_+)-Q(\scri^-_-)\,,
\end{split}
\end{equation}
the conservation law \eqref{BMS conservation} takes the form 
\begin{equation}
\label{BMS conservation 2}
Q(i^0)=Q(\scri^+_+)-F(\scri^+)=Q(\scri^-_-)+F(\scri^-)\,.
\end{equation}
Invariance of the $\mathcal{S}$-matrix under BMS symmetries can be stated as
\begin{equation}
\left[Q(i^0),\mathcal{S} \right]=0\,,
\end{equation}
or using \eqref{BMS conservation 2},
\begin{equation}
\label{Ward identity}
\outstate \left(F(\scri^+)-Q(\scri^+_+) \right) \mathcal{S}+\mathcal{S} \left( Q(\scri^-_-)+F(\scri^-) \right) \instate=0\,.
\end{equation}
One is left to evaluate the effect of inserting these charges and fluxes in the $\mathcal{S}$-matrix elements. We will focus on superrotations in this work, although the discussion also applied to supertranslations thus far. The hard part of the superrotation flux $F_{\mathcal{Y}}^{hard}(\scri^+)$ generates the symmetry transformations of the massless outgoing states, while the charge $Q_{\mathcal{Y}}(\scri^+_+)$ at the top corner of $\scri$ generates those of the massive outgoing states. A similar statement applies to incoming states. Together these terms account for the tree-level subleading soft factor $S^{(1)}_n$ in \eqref{soft_factors}. Indeed, insertion of the linear piece in the soft flux \eqref{superrotation fluxes}, namely
\begin{equation}
F^{soft,0}_{\mathcal{Y}}= -\frac{2}{\kappa^2} \int d^2z\, \mathcal{Y}\, \partial^3 \mathcal{N}^{(1)}_{\bar z \bar z}\,, 
\end{equation}
adds an external soft graviton to the original amplitude since the subleading soft news can be expressed in terms of graviton operators as \cite{Kapec:2014opa} 
\begin{equation}
\mathcal{N}^{(1)}_{\bar z \bar z}=\frac{i\kappa}{16\pi} \lim_{\omega \to 0} (1+\omega \partial_\omega)\, [a_-(\omega \hat q)-a_+^\dagger(\omega \hat q) ]\,.
\end{equation}
In this way one recovers the subleading soft graviton theorem at tree-level \cite{Kapec:2014opa,Campiglia:2015kxa}. In a previous publication \cite{Donnay:2022hkf} (see also \cite{Pasterski:2022djr}), it was shown that the infrared divergent one-loop corrections to the soft graviton theorem that were derived in \cite{Bern:2014oka} are accounted for by insertion of the remaining terms in the soft flux \eqref{superrotation fluxes}, which can be rewritten as 
\begin{equation}
\label{new soft terms}
F_{\mY}^{soft,new}=\frac{2}{\kappa^2}\int d^2 z\,  \mathcal{Y}  \left[-\partial^3 ( C^{(0)} \mathcal{N}^{(0)}_{\bar z \bar z} ) +3 \bar \partial^2\mathcal{N}^{(0)}_{zz} \partial C^{(0)}+ C^{(0)} \partial\bar\partial^2 \mathcal{N}_{zz}^{(0)}  \right]\,.
\end{equation}
The focus of that previous analysis was on recovering the result of Bern et al.~which concerns the IR-divergent one-loop contributions to the soft graviton theorem in case all particles are massless. Here we extend that analysis so that we recover the loop corrections found by Sahoo and Sen that also apply to the scattering of massive particles. As we discussed in section~\ref{section: log corrections} these results take the exact same form when expressed in terms of the quantities $\sigma_n\,, \hat \sigma'_{n+1}$ and $\hat S^{(0)}_n, S^{(1)}_{n}$, modulo the identification \eqref{epsilon log omega identification}. Hence it is just a matter of extending the derivation presented in \cite{Donnay:2022hkf} in way that it also applies to massive particles. 

The two terms in \eqref{log factor rewritten} will arise in two distinct ways. The first term $\hat \sigma'_{n+1}\, \hat S^{(0)}_n$ will be obtained by insertion of the soft flux contribution $F_{\mY}^{soft,new}$ into the $\mathcal{S}$-matrix, while the second term $S^{(1)J}_n \sigma_n$ will arise from a contribution to the charges at timelike infinity, $Q_{\mY}^{i^+} \equiv Q_{\mY}(\scri^+_+)$ and $Q_{\mY}^{i^-} \equiv Q_{\mY}(\scri^-_-)$ which is associated with the gravitational dressing of the massive scalar fields. It is interesting to note that the imaginary part of these terms can be given a very clear classical interpretation: $\text{Im}\, (\hat \sigma'_{n+1}\, \hat S^{(0)}_n)$ results from the gravitational drag exerted by the hard particles on the soft graviton, while $\text{Im}\, (S^{(1)J}_n \sigma_n)$ results from the long range `Coulombic'/Newtonian interaction between the hard particles. Quantum mechanically, $\hat \sigma'_{n+1}\, \hat S^{(0)}_n$ and $S^{(1)J}_n\, \sigma_n$ arise from Feynman diagrams where the soft graviton attaches to a virtual soft graviton or to an external hard particle, respectively \cite{Sahoo:2018lxl}.  

We start by inserting the soft flux contribution $F_{\mY}^{soft,new}$ as given in \eqref{new soft terms} into the $\mathcal{S}$-matrix, upon using the formulae \eqref{Goldstone insertion} and \eqref{N0 insertion} for the insertions of $C^{(0)}$ and $\mathcal{N}^{(0)}_{zz}$, respectively. This yields
\begin{equation}
\label{first terms}
\begin{split}
\outstate & F_{\mY}^{soft,new}(\scri^+)\, \mathcal{S}+\mathcal{S}F_{\mY}^{soft,new}(\scri^-)\instate\\
&=
-\frac{i\kappa}{8\pi\epsilon}  \int d^2z\,  \mY \left[ \partial^3 (\hat \sigma_{n+1}'\, \hat S^{(0)-}_n)  -\hat \sigma'_{n+1}\, \partial\bar \partial^2 \hat S^{(0)+}_n-3 \partial \hat \sigma'_{n+1} \, \bar \partial^2 \hat S^{(0)+}_n  \right]\outstate \mathcal{S} \instate\\
&=-\frac{i\kappa}{16\pi\epsilon}  \int d^2z\,  \mY\, \partial^3 (\hat \sigma_{n+1}'\, \hat S^{(0)-}_n)\,  \outstate \mathcal{S} \instate\,,
\end{split}
\end{equation}
where in the second equality we used (5.18) in \cite{Donnay:2022hkf}. This accounts for the first term in \eqref{log factor rewritten}.

We now move to the derivation of the remaining term, $S^{(1)J}_n \sigma_n$, from the hard part of the superrotation generators. This includes the fluxes acting on the massless particles at null infinity and the charges acting on massive particles at timelike infinity. As massive fields propagate towards future timelike infinity $i^+$ rather than null infinity, it is convenient to use hyperbolic coordinates $(\tau,x^a)$ within the future light cone,
\begin{equation}
ds^2=-d \tau^2+\tau^2\, h_{ab}\, dx^a\, dx^b=-d \tau^2+\tau^2 \left(\frac{d\rho^2}{\rho^2}+\rho^2\, dz\,d\bar z \right)\,.
\end{equation}
The three-dimensional hyperboloid $\mathcal{H}$ covered by the coordinates $x^a=(\rho,z,\bar z)$ describes the directions of approach to $i^+$. Obviously a similar construction applies in the neighborhood of past timelike infinity $i^-$. In this coordinate system superrotations are parametrised by a vector field $\mathcal{Y}_{\mathcal{H}}^a\, \partial_a$ satisfying $D_a \mathcal Y_{\mathcal H}^a=0$ and determined directly by the standard superrotation parameters $\mathcal{Y}^A=(\mathcal{Y},\bar{\mathcal{Y}})$ through the integral relation \cite{Campiglia:2015kxa,Campiglia:2015lxa}
\begin{equation}
\mathcal Y_{\mathcal H}^a(\hat p)= \int d^2 \hat q \, G_A^a(\hat p;\hat q) \mathcal Y^A(\hat q)\,, \qquad \hat p=\hat p(\rho,z,\bar z)\,, \quad \hat q=\hat q(w,\bar w)\,,
\end{equation}
where the `bulk-boundary propagator' $G^a_A$ admits the representation
\begin{equation}
\label{GA representation}
G^a_w(\hat p;\hat q) \partial_a=\frac{i}{4\pi}\, \partial_w^3 \left(\frac{\hat p^\mu\,  \varepsilon^-_{\mu\nu}(\hat q)\, \hat q_\lambda}{\hat p \cdot \hat q}\right) J^{\lambda\nu}(\hat p)\,.
\end{equation}
Asymptotically superrotations are diffeomorphisms generated by the vector field $\mathcal{Y}_{\mathcal{H}}^a \partial_a$ such that the charge generating the corresponding asymptotic transformation of the massive matter fields is simply given by \cite{Campiglia:2015kxa}
\begin{equation}
\label{charge i+}
Q_{\mathcal{Y}}^{i^+}=\lim_{\tau \to \infty}  \frac{\tau^3}{8}\int_{\mathcal{H}} d^3 \hat p\, \mathcal{Y}^a_{\mathcal{H}}(\hat p)\, T_{\tau a}(\tau,\hat p)\,,
\end{equation}
where $T_{\mu\nu}$ is the matter stress tensor and $d^3\hat p$ denotes the volume element of the unit hyperboloid,
\begin{equation}
\int d^3 \hat p\, E_{\hat p}\, \delta^3(\hat p)=1\,, \qquad E_{\hat p}\equiv \sqrt{|\vec p/m|^2+1}\,.
\end{equation}
Ideally one should prove that this quantity is indeed equal to $Q_{\mathcal{Y}}(\scri^+_+)$ as defined in \eqref{charge at corner}. Progress in this direction has been reported in \cite{Compere:2023qoa}, however here we will simply assume its validity as customarily done. 
For simplicity we will analyse the case of massive scalar fields which is appropriate to the scattering of massive spinless particles. Asymptotically free massive scalar fields can be written in terms of the creation/annihilation operators \cite{Campiglia:2015kxa}, 
\begin{equation}
\phi^{(free)} (\tau, x^a) = \frac{\sqrt{m}}{2(2\pi \tau)^{3/2}}\left( b(p)\, e^{-i\tau m} +b(p)^\dagger\, e^{i\tau m}  \right) +O(\tau^{-5/2}) \,,
\end{equation}
with the position-momentum relation $p= m\, \hat p(x^a)$ given as in \eqref{massive para}, following from a stationary phase approximation of the  momentum mode decomposition of the field in the large-$\tau$ limit. In turn the stress tensor components of interest take the form
\begin{equation}
\begin{split}
T_{\tau a}^{(free)}&=\frac{im^2}{4(2\pi \tau)^3} \left[b^\dagger \partial_a b-\partial_a b^\dagger b+ \frac{1}{2} \partial_a\left( b^{\dagger 2} e^{2im \tau}-b^2 e^{-2im \tau}\right)\right]+O(\tau^{-4})\,,
\end{split}
\end{equation}
such that using $D_a \mathcal Y_{\mathcal H}^a=0$ the charge \eqref{charge i+} becomes
\begin{equation}
Q_{\mathcal{Y}}^{i^+(free)}=  \frac{im^2}{32(2\pi)^3} \int_{\mathcal{H}} d^3 \hat p\, \mathcal{Y}^a_{\mathcal{H}}(\hat p) \left( b^\dagger \partial_a b-\partial_a b^\dagger b\right)\,.
\end{equation}

However the charge formula \eqref{charge i+} really features matter fields in gravitational interaction rather than free fields. This will be taken care of by dressing the free field operators. In particular the `classical' part of the dressing will properly account for the long-range gravitational interaction between scattered particles. 
Dressing the field $\phi$ hence amounts to performing the replacement 
\begin{equation}
b(\hat p)^\dagger \quad \mapsto \quad \tilde b(\hat p)^\dagger=\exp \left[-\frac{i m}{2} \int d^2 \hat q\, \mathcal{G}(\hat p;\hat q)\, C^{(0)}(\hat q) \right] b(\hat p)^\dagger\,,
\end{equation}
as can be directly inferred from \eqref{factorisation operators}-\eqref{massive W}. Indeed in the language of celestial holography the creation operator $b(p)^\dagger$ is represented by a local operator $\mathcal O(p)$.
The dressed superrotation charge thus acquires an extra contribution,
\begin{equation}
Q_{\mathcal{Y}}^{i^+}=Q_{\mathcal{Y}}^{i^+(free)}-\Delta Q_{\mathcal{Y}}^{i^+}\,,
\end{equation}
where 
\begin{equation}
\Delta Q_{\mathcal{Y}}^{i^+}=\frac{m^3}{32(2\pi)^3}\int d^2  \hat q \int d^3 \hat p\, \mathcal{Y}^a_{\mathcal{H}}(\hat p)\,  \partial_a \mathcal{G}(\hat p;\hat q)\, C^{(0)}(\hat q)\,  b(\hat p)^\dagger b(\hat p)\,.
\end{equation}

The contribution of the free charge $Q_{\mathcal{Y}}^{i^+(free)}$ to the Ward identity \eqref{Ward identity} yields the leading soft factor $\hat S^{(0)}_n$ \cite{Campiglia:2015kxa}. We are left to evaluate the result of inserting $\Delta Q_{\mathcal{Y}}^{i^+}$ and $\Delta Q_{\mathcal{Y}}^{i^-}$ into $\mathcal{S}$-matrix elements. We will set $\bar{\mathcal{Y}}=0$ for ease of notation. First we recall the identities \cite{Himwich:2020rro,Donnay:2022hkf}
\begin{align}
\sigma_n &= -\frac{\epsilon}{8\kappa^2} \sum_{ij=1}^n  m_i m_j\int d^2 \hat q\, d^2 \hat q'\, \mathcal{G}(\hat p_i;\hat q) \mathcal{G}(\hat p_j;\hat q') \langle C^{(0)}(\hat q) C^{(0)}(\hat q') \rangle\,,
\end{align}
\begin{equation}
\label{Sigma prime convolution}
\begin{split}
\hat \sigma'_{n+1}(\hat q')&=-\frac{\epsilon}{2\kappa^2} \sum_{i=1}^n m_i \int d^2 \hat q\, \mathcal{G}(\hat p_i;\hat q)\, \langle C^{(0)}(\hat q') C^{(0)}(\hat q) \rangle\\
&=\frac{1}{(8\pi)^2} \sum_{i=1}^n m_i \int d^2 \hat q\, \mathcal{G}(\hat p_i;\hat q)\,  (\hat q \cdot \hat q') \ln (\hat q \cdot \hat q')\,,
\end{split}
\end{equation}
such that
\begin{align}
\label{d3 S Sigma}
\partial^3 S^{(1)J-}_n(\hat q)\, \sigma_n &=\frac{1}{2} \sum_{i}^n m_i \int d^2 \hat q'\, \mathcal{G}(\hat p_i;\hat q')\, \partial^3 S^{(1)J-}_n(\hat q)\, \hat \sigma'_{n+1}(\hat q')\,.
\end{align}
We then have 
\begin{equation}
\label{Delta Q insertion}
\begin{split}
\outstate &\Delta Q_{\mathcal{Y}}^{i^+} \mathcal{S} - \mathcal{S} \Delta Q_{\mathcal{Y}}^{i^-}\instate\\
&=  \frac{1}{16}\sum_{i} m_i \int d^2  \hat q \,    \mathcal{Y}^a_{\mathcal{H}}(\hat p_i)\partial_a\mathcal{G}(\hat p_i;\hat q)\,\outstate   C^{(0)}(\hat q) \mathcal{S}\instate\\
&= \frac{-i \kappa^2}{16\epsilon} \sum_{i} m_i \int d^2  \hat q \, \mathcal{Y}^a_{\mathcal{H}}(\hat p_i)  \partial_a \mathcal{G}(\hat p_i;\hat q)\, \hat \sigma'_{n+1}(\hat q)\,  \outstate \mathcal{S}\instate\\
&= \frac{i}{32} \sum_{ij} m_i m_j\, \int d^2  \hat q'\, \mathcal{G}(\hat p_j;\hat q')  \int d^2 \hat q\,   \mathcal{Y}^a_{\mathcal{H}}(\hat p_i)  \partial_a\mathcal{G}(\hat p_i;\hat q) \, \langle C^{(0)}(\hat q) C^{(0)}(\hat q') \rangle\,  \outstate \mathcal{S}\instate\\
&=\frac{-i\kappa}{32\pi\epsilon}\sum_{j} m_j \int d^2  \hat q'\, \mathcal{G}(\hat p_j;\hat q')  \int d^2  \hat q \, \mathcal{Y}(\hat q)\, \partial^3 S^{(1)J-}_n(\hat q)\, \hat \sigma'_{n+1}(\hat q')\,  \outstate \mathcal{S}\instate\\
&=\frac{-i\kappa}{16\pi\epsilon} \int d^2  \hat q \, \mathcal{Y}(\hat q)\, \partial^3 S^{(1)J-}_n(\hat q)\, \sigma_n\,  \outstate \mathcal{S}\instate\,.
\end{split}
\end{equation}
In the first equality we acted on the massive external states with the particle number operator \cite{Campiglia:2015kxa}
\begin{equation}
b(\hat p)^\dagger\, b(\hat p)\, |p' \rangle=m^{-2} (2\pi)^3\, (2E_{\hat p})\, \delta^3(\hat p-\hat p')\,|p' \rangle\,. 
\end{equation}
The second equality is obtained from \eqref{Goldstone insertion}, while the third equality relies on \eqref{Sigma prime convolution}. The fourth equality holds thanks to \eqref{GA representation} together with \eqref{Sigma prime convolution}. The last equality follows directly from \eqref{d3 S Sigma}.

We observe that the last expression of \eqref{Delta Q insertion} indeed accounts for the second term $S^{(1)J}_n \sigma_n$ in \eqref{log factor rewritten}. Together with \eqref{first terms} we have therefore recovered all logarithmic corrections to the soft graviton theorems when external states are massive. The treatment of massless particles goes along similar lines. In particular it requires dressing the massless fields (including the graviton field if some of the scattered particles are hard gravitons) in the hard part of the flux $F^{hard}_{\mathcal{Y}}$ given in \eqref{superrotation fluxes}, which should then produce the corresponding massless contributions in $S^{(1)J}_n \sigma_n$. The massless contributions to $\hat \sigma'_{n+1}\, \hat S^{(0)}_n$ are still captured by \eqref{first terms} without modifications. Note that the result for massless particles is also directly obtained taking the corresponding massless limits.

\subsection*{Acknowledgements}  We thank Francesco Alessio, Shamik Banerjee, Miguel Campiglia, Geoffrey Compère, Laurent Freidel, Alok Laddha, Andrea Puhm and Ali Seraj for stimulating discussions. S.A.~and L.D.~are partially supported by INFN Iniziativa Specifica ST\&FI. K.N.~is supported by the STFC grants ST/P000258/1 and ST/T000759/1. R.R.~is supported by the Titchmarsh Research Fellowship in Mathematical Physics at the University of Oxford.  

\subsection*{References}
\bibliographystyle{style}
\renewcommand\refname{\vskip -1.3cm}
\bibliography{references}
\end{document}